\documentclass[twocolumn]{aastex62}

\usepackage{natbib}
\usepackage{amsmath}
\usepackage{subfigure}
\setcitestyle{sort,comma}
\usepackage{csvsimple}

\graphicspath{{./}{figures/}}

\shorttitle{Water Vapor and Clouds on the Habitable-Zone Sub-Neptune Exoplanet K2-18}
\shortauthors{Benneke et al.}

\begin{document}

\title{\large{Water Vapor and Clouds on the Habitable-Zone Sub-Neptune Exoplanet K2-18b}}

\correspondingauthor{Bj\"{o}rn Benneke}
\email{bbenneke@astro.umontreal.ca}

\author{Bj\"{o}rn Benneke} 
\affil{Department of Physics and Institute for Research on Exoplanets, Universit\'{e} de Montr\'{e}al, Montreal, QC, Canada}

\author{Ian Wong}
\affil{Department of Earth, Atmospheric, and Planetary Sciences, Massachusetts Institute of Technology, 77 Massachusetts Ave, Cambridge, MA, 02139, USA}
\affil{51 Pegasi b Fellow}

\author{Caroline Piaulet} 
\affil{Department of Physics and Institute for Research on Exoplanets, Universit\'{e} de Montr\'{e}al, Montreal, QC, Canada}

\author{Heather A. Knutson} 
\affil{Division of Geological and Planetary Sciences, California Institute of Technology, Pasadena, CA 91125, USA}

\author{Joshua Lothringer} 
\affil{Department of Physics and Astronomy, Johns Hopkins University, Baltimore, MD 21218, USA}

\author{Caroline V. Morley}
\affil{Department of Astronomy, University of Texas, Austin, TX 78712, USA}

\author{Ian J.M. Crossfield}
\affil{Department of Physics and Kavli Institute of Astronomy, Massachusetts Institute of Technology, 77 Massachusetts Ave, Cambridge, MA, 02139, USA}

\author{Peter Gao}
\affil{Department of Astronomy, University of California - Berkeley , Berkeley, CA, 94720, USA}
\affil{51 Pegasi b Fellow}

\author{Thomas P. Greene}
\affil{NASA Ames Research Center, Moffett Field, CA, 94035, USA}

\author{Courtney Dressing}
\affil{Department of Astronomy, University of California - Berkeley, Berkeley, CA, 94720, USA}


\author{Diana Dragomir}
\affiliation{Department of Physics and Astronomy, University of New Mexico, Albuquerque, NM, USA}

\author{Andrew W. Howard}
\affiliation{Department of Astronomy, California Institute of Technology, Pasadena, CA 91125, USA}


\author{Peter R. McCullough}
\affiliation{Department of Physics and Astronomy, Johns Hopkins University, Baltimore, MD 21218, USA}

\author{Eliza M.-R. Kempton}
\affiliation{Department of Astronomy, University of Maryland, College Park, MD 20742, USA}
\affiliation{Department of Physics, Grinnell College, 1116 8th Avenue, Grinnell, IA 50112, USA}

\author{Jonathan J. Fortney}
\affiliation{Department of Astronomy, University of California, Santa Cruz, CA 95064, USA}

\author{Jonathan Fraine}
\affiliation{Center for Extrasolar Planetary Systems, Space Science Institute, Boulder, CO 80301, USA}







\begin{abstract}

Results from the \textit{Kepler} mission indicate that the occurrence rate of small planets ($<3\,R_\oplus$) in the habitable zone of nearby low-mass stars may be as high as 80\%. Despite this abundance, probing the conditions and atmospheric properties on any habitable-zone planet is extremely difficult and has remained elusive to date. Here, we report the detection of water vapor and the likely presence of liquid and icy water clouds in the atmosphere of the $2.6\,R_\oplus$ habitable-zone planet K2-18b. The simultaneous detection of water vapor and clouds in the mid-atmosphere of K2-18b is particularly intriguing because K2-18b receives virtually the same amount of total insolation from its host star ($1368_{-107}^{+114}$\,W\,m$^{-2}$) as the Earth receives from the Sun (1361\,W\,m$^{-2}$), resulting in the right conditions for water vapor to condense and explain the detected clouds. In this study we observed nine transits of K2-18b using \textit{Hubble Space Telescope}/WFC3 in order to achieve the necessary sensitivity to detect the water vapor, and we supplement this data set with \textit{Spitzer} and \textit{K2} observations to obtain a broader wavelength coverage. While the thick hydrogen-dominated envelope we detect on K2-18b means that the planet is not a true Earth analog, our observations demonstrate that low-mass habitable-zone planets with the right conditions for liquid water are accessible with state-of-the-art telescopes. 
\end{abstract}



\section{Introduction} \label{sec:intro}

The recent discovery of the transiting $8.63\pm1.35$~M$_\oplus$ exoplanet K2-18b in the habitable zone of a bright, nearby M3-dwarf provides us with an opportunity to carry out a spectroscopic study of the atmosphere of a habitable-zone planet outside our solar system \citep{montet_stellar_2015,benneke_spitzer_2017,cloutier_characterization_2017,sarkis_carmenes_2018,cloutier_confirmation_2019}. K2-18b is an intriguing planet because its equilibrium temperature ($255\pm4$~K at an albedo of $A=0.3$) is potentially very close to that of the Earth (257~K). The planet's predicted temperature provides the right conditions for water vapor to condense to the liquid phase in its atmosphere. While K2-18b has a much shorter orbital period (33 days) than the Earth, its host star is also cooler and smaller than the Sun (3457 K, 0.45 R$_\odot$) resulting in only 2.53\% of the Sun's luminosity. As a result, the total insolation received by K2-18b ($1368_{-107}^{+114}$~W/m$^2$) is very close to that received by the Earth (1361 W/m$^2$), putting K2-18b firmly in the habitable zone.

Compared to habitable-zone planets around Sun-like stars, habitable-zone planets around M dwarfs offer two key advantages for atmosphere studies \citep{nutzman_design_2008}. The small diameter of the star results in larger transit signatures as the amplitude of transit and the atmospheric signal scale inversely with the square of the stellar radius. Furthermore, the short orbital periods for habitable-zone planets around mid M dwarfs (30--70~days) enables the observation of repeated transits within a relatively short time frame. This means that for K2-18b, we were able to observe nine transits with Wide Field Camera 3 on the \textit{Hubble Space Telescope} (\textit{HST}/WFC3) within a period of 3~yr.

\textit{Kepler} showed that as much as 80\% of M dwarfs host small planets ($<3 R_\oplus$) in the habitable zone \citep{dressing_occurrence_2013,dressing_occurrence_2015,kopparapu_revised_2013,silburt_statistical_2015,farr_counting_2015}, suggesting that planets like K2-18b may be common. However, most of the M dwarf planetary systems detected in the original \textit{Kepler} survey are extremely faint, making spectroscopic characterization of these planets prohibitively inefficient. Fortunately, K2-18b orbits a relatively bright ($K=8.89$) host star, permitting detailed characterization of its atmosphere.

Beyond its potentially temperate climate, K2-18b also occupies an interesting niche in mass-radius space. Very little is currently known about the bulk and atmospheric compositions of planets with masses between those of Earth and Neptune. These planets have no analogs in the solar system, and aside from the recent atmospheric detection for GJ~3470b \citep{benneke_sub-neptune_2019}, most atmospheric studies resulted in nondetections due to the prevalence of high-altitude clouds \citep{knutson_hubble_2014,kreidberg_clouds_2014,crossfield_two_2017}. Refined population studies of the radius distribution of sub-Neptune-sized planets have revealed a significant drop in the planet occurrence rates between near 1.5--2.0~$R_\Earth$ \citep{fulton_california-keplersurvey._2017, fulton_california-kepler_2018}. Photoevaporation \citep{owen_kepler_2013,lopez_role_2013,lopez_how_2018} and core-powered mass loss \citep{schlichting_formation_2018} are two theories that can reproduce this observed drop in the occurrence rate. According to photoevaporation models, the most highly irradiated super-Earths are expected to have primarily rocky compositions and relatively small radii, while less irradiated super-Earths are able to retain a modest (few percent in mass) primordial hydrogen-rich atmosphere that inflates their observed radii to values greater than $2.0~R_\Earth$. With a relatively low incident flux and a measured radius of $2.61~R_\Earth$, K2-18b would then be expected to host an extended hydrogen-rich atmosphere, making it a favorable target for atmospheric characterization studies using the transmission spectroscopy technique.

\begin{table}
\centering
\begin{tabular}{ c   c c c}
\hline
\hline
Instrument & Wavelength ($\mu$m) & UT Start Date \\
\hline
\textit{Kepler}/\textit{K2} & 0.43 -- 0.89 & 2014 Nov 14\\
&& \\
\textit{HST}/WFC3 G141 & 1.05 -- 1.70  & 2015 Dec 06\\
&& 2016 Mar 14\\
&& 2016 May 19\\
&& 2016 Dec 02\\
&& 2017 Jan 04\\
&& 2017 Feb 06\\
&& 2017 Apr 13\\
&& 2017 Nov 30\\
&& 2018 May 13\\
\\
\textit{Spitzer}/IRAC Ch1 & 3.15 -- 3.94 & 2016 Mar 14\\
&& 2016 Aug 26\\
\\
\textit{Spitzer}/IRAC Ch2 & 3.96 -- 5.02 & 2015 Aug 29\\
\hline
\hline
\end{tabular}
\caption{\label{tab:obs} Summary of all Transit Observations of K2-18b Obtained for This Study.}
\end{table}

In this work, we present the detection of water vapor and clouds in the atmosphere of the habitable-zone exoplanet K2-18b. In Section 2 we describe our observations obtained with \textit{HST}, \textit{Spitzer}, and \textit{K2}, as well as the techniques used to reduce the data and produce spectrophotometric lightcurves. In Section 3 we discuss the data analysis and present the best-fit white light-curve parameters and transmission spectrum. Importantly, we also provide a refined stellar characterization. Our atmospheric modeling analysis is described in Section 4. The main results are presented in Section 5, and the possibility of liquid water clouds is discussed in Section 6.

\section{Observations and data reduction}\label{sec:obs_red}

Our team observed the transiting habitable-zone exoplanet K2-18b with Wide Field Camera 3 on the \textit{Hubble Space Telescope} as part of two large spectral surveys of low-mass exoplanets (GO 13665 and GO 14682; PI Benneke). Building up sufficient signal-to-noise was possible for this habitable-zone planet because the shorter orbital period (33 days) enabled us to observe nine independent transits within 3~yr. The SNR of the transmission signal was further boosted by the small stellar radius, which amplifies the signal of the planet and atmosphere during transit. 
We complement the \textit{HST}/WFC3 observations with two new \textit{Spitzer} transit observations taken at 3.6 $\mu$m (Program 12081, PI Benneke) as well as previously published \textit{Spitzer} (4.5 $\mu$m) and \textit{Kepler/K2} transit observations from \citet[][see our Table \ref{tab:obs}]{benneke_spitzer_2017}. 

\subsection{\textnormal{HST}/WFC3 transits}

Each of our nine \textit{HST}/WFC3 visits spanned 6.5 hours and consisted of four full telescope orbits separated by 45~minute gaps in data collection due to Earth occultation. We obtained the \textit{HST}/WFC3 time series with the G141 grism in spatial scan mode. In this configuration the telescope is scanned during the exposure, moving the stellar spectrum across the detector perpendicular to the dispersion direction  \citep{deming_infrared_2013,kreidberg_clouds_2014}.  This allows for a significantly higher efficiency when observing bright stars like K2-18b. In order to minimize instrumental overheads we utilized both forward and backward scans with the maximum possible duration, covering a large fraction of the $256 \times 256$ pixel detector subarray used for fast readouts. We do not include the final HST/WFC3 transit in the analysis because it was corrupted by telescope guiding errors which resulted in the spectrum migrating off the detector subarray preventing us from doing useful science with the this particular transit observations.

Following standard procedure \citep[e.g.,][]{deming_infrared_2013}, we minimize the contribution from the sky background by subtracting consecutive nondestructive reads and then coadding these background-subtracted subexposures. We then use the wavelength-dependent flat-field data provided by STScI to build flat-fielded images. Bad pixels are removed and replaced by the corresponding value in a normalized row-added flux template.

The combined effect of the spatial scans and the position-dependent grism dispersion results in a slightly trapezoidal shape for the illuminated patch. The trapezoidal shape is due to a small difference in the dispersion on the detector along its y-axis, yielding a systematic horizontal shift of each wavelength by 2--3 pixels across the scan. To correctly capture this effect, we integrate over trapezoidal wavelength bins instead of rectangular ones, built from lines of constant wavelength obtained from our 2D wavelength solution computed across the detector. We use the same procedure as in \citet{benneke_sub-neptune_2019} for the flux integration, avoiding any presmoothing and accounting for the partial pixel flux along the bin boundaries to ensure total flux conservation. We also account for small $x$ position shifts in each frame to correct for the small observed drift in the star's position across the observations.

\subsection{\textnormal{Spitzer}/IRAC transits}

We obtained two new transit observations of K2-18b with \textit{Spitzer} at 3.6~$\mu$m (Program 12081, PI Benneke) and reanalyzed the 4.5~$\mu$m transit observation previously published in \citet{benneke_spitzer_2017}. The 3.6~$\mu$m transit observations were preceded by 30~minute preobservations in peak-up mode to mitigate telescope drift and temperature variations associated with a recent shift in pointing \citep{grillmair_pointing_2012}. We used 0.4\,s exposures and observed for a total of $\sim$8\,hr on 2016 March 14 and 2016 August 26.   

We follow standard procedure for \textit{Spitzer}/IRAC image processing and start from the flat-fielded and dark-subtracted ``Basic Calibrated Data'' (BCD) images. We use the method presented in \citet{kammer_spitzer_2015} for background estimation, and determine the position of the star following  \citet{benneke_sub-neptune_2019}. We then chose the aperture, trim duration, and bin size to minimize both the RMS of the unbinned residuals and the time-correlated noise in the data for individual fits.

\subsection{\textnormal{Kepler/K2} transits}

We supplement our dataset with two previously published \textit{K2} transits of K2-18b, corrected for variations associated with telescope jitter and cosmic ray hits as described in \citet{benneke_spitzer_2017}.

\begin{table*}
\centering
\begin{tabular*}{\textwidth}{l @{\extracolsep{\fill}} ccl}
\hline
\hline
Parameter & Unit & Value & Comment \\
\hline

\multicolumn{3}{l}{\rule{0pt}{4ex}\textit{Host star:}}\\
$M_\star$ & $M_\odot$ & $0.4951\pm0.0043$ & \citet{cloutier_confirmation_2019}\\
$R_\star$ & $R_\odot$ & $0.4445\pm0.0148$ & this paper\\
$T_{\star,\mathrm{eff}}$ & $M_\odot$ & $3457\pm39$ & \citet{benneke_spitzer_2017}\\
$L_{\star}$ & $L_\odot$ & $0.0253\pm0.0021$ & this paper \\
Distance & pc & $38.025\pm0.079$ & \citet{cloutier_confirmation_2019}\\

\multicolumn{3}{l}{\rule{0pt}{4ex}\textit{Orbit:}}\\
$R_P/R_\star$ & -- & $0.05375_{-0.00012}^{+0.00012}$  & this paper \\
$T_0$ & BJD$_{\mathrm{UTC}}$ & $2457725.551189_{-0.000068}^{+0.000067}$  & this paper \\
$P$ & days & $32.940045_{-0.000010}^{+0.000010}$  & this paper \\
$a_p$ & au & $0.15910_{-0.00047}^{+0.00046}$ & this paper \\
$b$ & -- & $0.641_{-0.015}^{+0.015}$  & this paper \\

\multicolumn{3}{l}{\rule{0pt}{4ex}\textit{Planet:}}\\
$R_p$ & $R_\oplus$ & $2.610\pm0.087$  & this paper \\
$M_p$ & $M_\oplus$ & $8.63\pm1.35$ & \citet{cloutier_confirmation_2019}\\
$\rho_p$ & $\mathrm{g\,cm^{-3}}$ & $2.67_{-0.47}^{+0.52}$ & this paper \\ 
$g_p$ & W\,m$^{-2}$ & $12.43_{-2.07}^{+2.17}$ & at 10~mbar, this paper \\ 
$S_p$ & $S_\oplus$ & $1.005_{-0.079}^{+0.084}$ & this paper, $1368_{-107}^{+114}\,\mathrm{W/m^2}$\\
$T_{\mathrm{eq}}$ & K & $254.9\pm3.9$ & this paper, for $A_B=0.3$\\

\multicolumn{3}{l}{\rule{0pt}{4ex}\textit{Atmosphere:}}\\
$\log \mathrm{H_2O}$ & -- & $-2.08_{-1.39}^{+1.03}$ & $0.0337-8.88\%$\\ 
$\log \mathrm{(H_2O / H_2O_{solar})}$ & -- & $1.18_{-1.39}^{+1.03}$ & $0.615-162$ $\times$ solar\\
$\log p_\mathrm{cloud-top}$ & mbar & $1.37_{-0.48}^{+0.77}$ & $7.74-139$ mbar\\ 
$\mu_{\mathrm{atm}}$ & -- & $2.42_{-0.12}^{+1.27}$\\

\hline
\hline
\end{tabular*}
\caption{\label{tab:orb_para} Parameters of the Host Star K2-18, the Planet K2-18b, and Its Atmosphere.}
\end{table*}

\section{Data Analysis}

\subsection{Updated stellar parameters}

For the correct analysis and interpretation of the K2-18b observations, it is absolutely crucial to update the stellar and planetary bulk parameters compared to the discovery and confirmation papers \citep{montet_stellar_2015,benneke_spitzer_2017}. This is particularly important for K2-18b because \textit{GAIA} DR2 substantially updated the parallax distance from the previously estimated 34 to $38.025\pm0.079$~pc \citep[see also][]{cloutier_confirmation_2019}. In its final consequence, this results in a substantial update to the planet radius from the previous value of $R_p=2.29\, R_\oplus$ to $R_p=2.61\,R_\oplus$. Not correcting the stellar radius led \citet{tsiaras_water_2019} to use a planetary density that was 27\% too high and a surface gravity that was 18\% too high. The overestimated planetary density may have led to the continued classification of K2-18b as a super-Earth, even though the radius of $R_p=2.61\,R_\oplus$ puts K2-18b directly in the center of the intriguing sub-Neptune population \citep{fulton_california-kepler_2018}. Besides, the overestimated surface gravity unavoidably resulted in a systematic offset in the atmospheric scale height for each model in the atmospheric retrieval. 

To revise the stellar parameters, we first calculate the distance modulus from the \textit{GAIA} DR2 stellar parallax to be $\mu=2.900\pm0.005$, where we add $30\,\mathrm{\mu as}$ systematic offset in the measured parallax \citep{cloutier_confirmation_2019}. By propagating the uncertainties in K2-18's $K$-band magnitude ($K=8.899\pm0.019$), we then find an absolute $K$-band magnitude of $M_K= 5.999\pm0.020$. Using this absolute $K$-band magnitude, we infer the stellar radius of K2-18b from the empirical M star correlation \citep{mann_how_2015} and the stellar mass from the empirical M dwarf mass-luminosity relation by \citet[][see our Table \ref{tab:orb_para}]{benedict_solar_2016}. We choose this approach for the stellar radius because it does not rely on the empirical mass-radius relationship, which we agreed with Dr. Cloutier to be the preferred approach. Either way the stellar radii presented here and the one from \citet{cloutier_confirmation_2019} are consistent at the 1$\sigma$. However, both estimates present a substantial update to the old pre-\textit{GAIA} stellar radius estimate, which led \citet{tsiaras_water_2019} to use an overestimated planetary density and surface gravity for their analyses. Finally, we use the effective stellar temperature from \citet{benneke_spitzer_2017} to compute the stellar luminosity as well as the insolation of the planet (Table \ref{tab:orb_para}).

\subsection{Transit White-light-curve Fitting} \label{sec:tr_fit}

\begin{figure}[t!]
\begin{center}
\includegraphics[width=\linewidth]{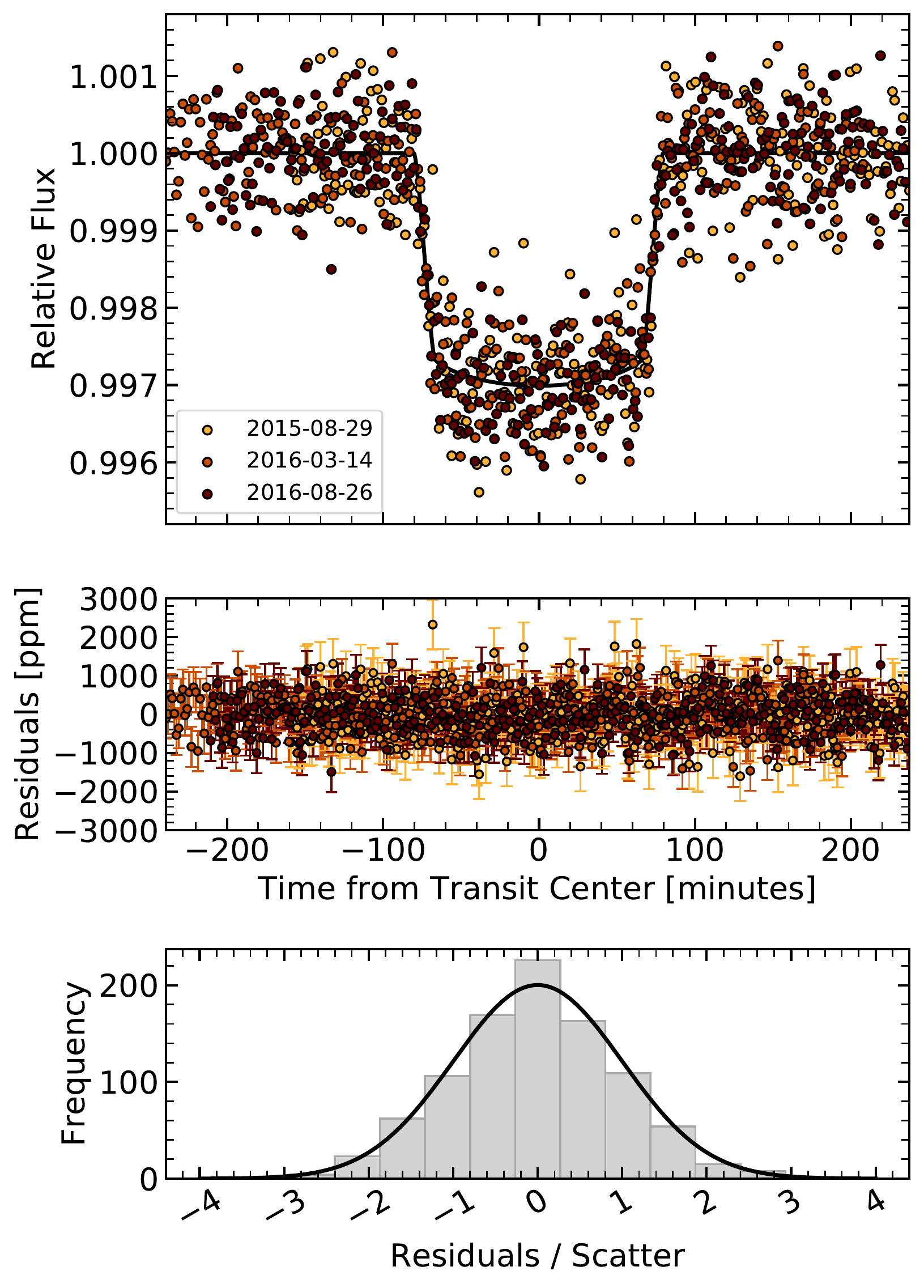}
\end{center}
\caption{Broadband light-curve fit from the joint analysis of the three \textit{Spitzer}/IRAC transit observations of K2-18b. The top panel shows the best-fitting model light curves (black curve), overlaid with the systematics-corrected data (circles). Residuals from the light-curve fits are shown in the middle panels. 
The bottom panels shows a histogram of the residuals normalized by the fitted photometric scatter parameter for each respective transit. The residuals follow the expected Gaussian distribution for photon noise limited observations.}
\label{fig:spitzer}
\end{figure}

We carry out a global analysis of our \textit{HST}/WFC3, \textit{K2}, and \textit{Spitzer} transit light curves within the ExoTEP analysis framework \citep{benneke_spitzer_2017, benneke_sub-neptune_2019}. ExoTEP jointly fits the transit and systematics models of all transits along with photometric noise parameters using a Markov Chain Monte Carlo (MCMC) method. The main astrophysical outputs of the white-light-curve analysis are the global transit parameters ($a/R_\star$, $b$), the transit ephemeris ($T_0$, $P$), and the transit depths in the \textit{K2}, \textit{HST}/WFC3, and \textit{Spitzer}/IRAC bandpasses (Table \ref{tab:orb_para}). Prior to the global MCMC fit, we analyze each transit light curve individually and then initialize the corresponding systematics parameters in the global fit at their best-fit values to ease convergence. We find that the planet-to-star radius ratio estimates in the same band are consistent within their statistical uncertainties across all epochs. The white-light-curve fits to the \textit{HST}/WFC3 and \textit{Spitzer} transits are depicted in Figures \ref{fig:HST} and \ref{fig:spitzer}.

\subsubsection{\textit{HST}/WFC3 Instrument Model}

We correct for systematic trends in the uncorrected WFC3 transit light curves by simultaneously fitting an analytical model-ramp function along with the astrophysical transit model. Following previous studies \citep[e.g.,][]{berta_flat_2012,deming_infrared_2013,kreidberg_clouds_2014}, we account for the possible presence of both visit-long slopes and orbit-long exponential ramps using the parametric instrumental systematics model:
\begin{equation}
S_{\mathrm{WFC3}} (t)=(c\,S(t)+v\,t_v ) \times (1 - \exp(-a\,t_{\mathrm{orb}}-b)).
\end{equation}
\noindent Here, $c$ is a normalization constant, $v$ is the visit-long linear slope, $a$ and $b$ describe the rate and amplitude of the orbit-long exponential slope, $S(t)$ is set to 1 for forward scans and $s$ for backward scans to account for the systematic flux offset between the scan directions, and $t_v$ and $t_{\mathrm{orb}}$ are the time in hours since the start of the visit and the start of the observations within the current orbit. Following standard procedure, we discard the first \textit{HST} orbit of each visit, as it systematically exhibits a stronger ramp than the three subsequent orbits. We also remove the first forward and backward scan exposures of each orbit.

\subsubsection{\textit{Spitzer}/IRAC Instrument Model}

We account for the presence of systematic variations in the \textit{Spitzer} time series due to subpixel inhomogeneities in the detector response using the pixel-level decorrelation (PLD) model \citep{deming_spitzer_2015,benneke_spitzer_2017}. For each \textit{Spitzer} visit, the systematics model
\begin{equation}
    S_{\mathrm{Spitzer}} (t_i)=1+A\,e^{-t_i/\tau}+m\,t+\frac{\sum_{k=1}^9 w_k D_k(t_i)}{\sum_{k=1}^9 D_k(t_i)}. 
\end{equation}
\noindent includes both a linear-exponential ramp in time ($A\,e^{-t_i/\tau}+m\,t$) and the PLD term. In this analytical model, $D_k (t_i )$ are the registered counts in each of the $3 \times 3$ pixels covering the central region of the point spread function. The coefficients $w_k$ are time-independent PLD weights. The 10 parameters in this model are fitted along with the transit model for all \textit{Spitzer} data sets.

\subsubsection{\textnormal{Kepler/K2} Instrument Model}

We use the detrended \textit{K2} light curves of K2-18b from \citet{benneke_spitzer_2017} and additionally fit a linear trend with time to allow for a residual slope in each visit superimposed with the transit signal.

\subsubsection{Transit Model and MCMC Analysis}

\begin{figure*}[!tbp]
  \centering
  \subfigure{\includegraphics[width=0.48\textwidth]{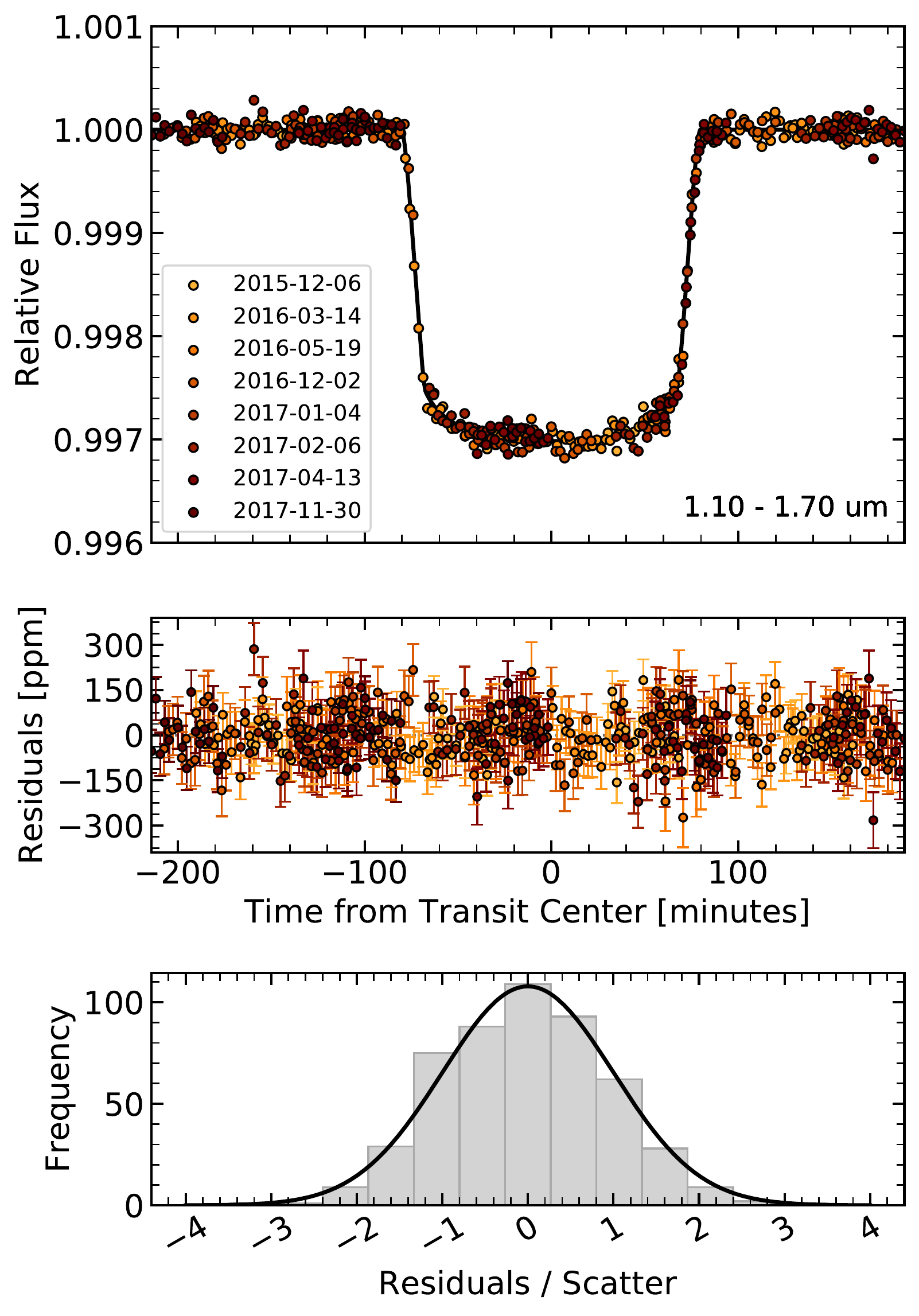}}
  \hfill
  \subfigure{\includegraphics[width=0.48\textwidth]{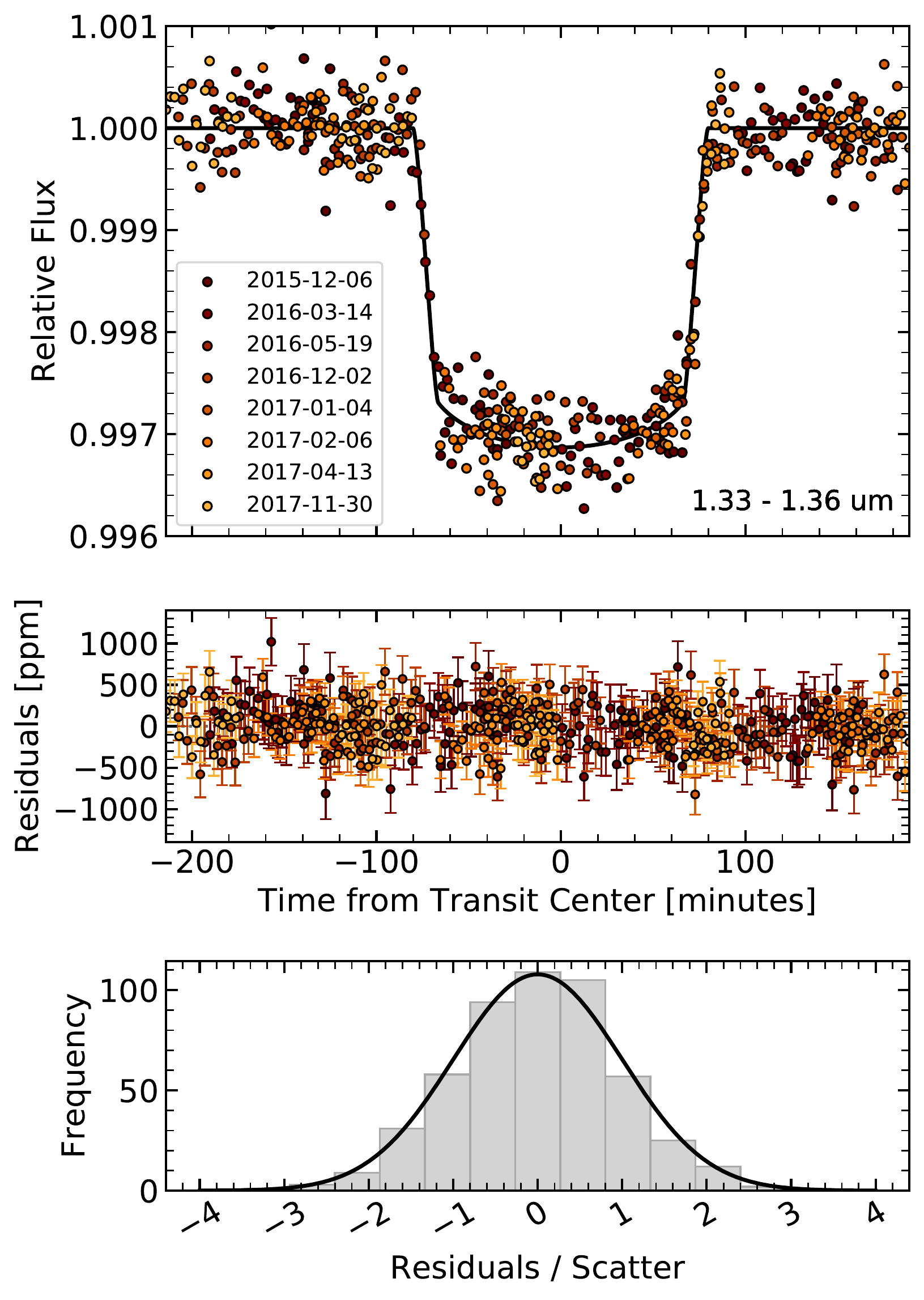}}
  \caption{White-light-curve fit (left) and a typical spectral-light-curve fit (right) from the joint analysis of the eight WFC3 transit observations of K2-18b. The top panel shows the best-fitting model light curves (black curve), overlaid with the systematics-corrected data (circles). Residuals from the light-curve fits are shown in the middle panels. The bottom panels shows a histogram of the residuals normalized by the fitted photometric scatter parameter for each respective transit. The residuals follow the expected Gaussian distribution for photon noise limited observations.}
  \label{fig:HST}
\end{figure*}

Our astrophysical transit light-curve model $f(t_i)$ is computed using the Batman module \citep{kreidberg_batman:_2015}. In the joint white-light-curve fit, four distinct transit depths are fitted for the four bandpasses of \textit{K2}, \textit{HST}/WFC3, \textit{Spitzer}/IRAC Channel 1, and \textit{Spitzer}/IRAC Channel 2. The orbital parameters are assumed to be consistent for all 12 transit light curves. We also fit limb-darkening coefficients using a quadratic law because the eight \textit{HST}/WFC3 combined provide full time coverage of the transit light curve including ingress and egress (Figure \ref{fig:HST}). The cadence of the observations is accounted for by integrating the model over time within each exposure. Finally, the total log-likelihood for a single set of parameters is calculated using
\begin{equation}
\begin{split}
\ln \mathcal{L} = \sum_{V=1}^N &-n_V \ln \sigma_V -\frac{n_V}{2} \ln 2\pi \\
&- \sum_{i=1}^{n_V}\frac{\left[d_V(t_i)-S_V(t_i)\times f_V(t_i)\right]^2}{\sigma_V^2},
\end{split}
\end{equation}
\noindent where $N$ is the number of visits, $n_V$ is the number of data points $d_V (t_i)$ in visit $V$, and $\sigma_V$ is a photometric noise parameter associated with each visit, simultaneously fitted to account for the possibility of variations in the scatter between independent visits. Each visit has a different systematics model $S_V (t_i)$ with additional free parameters specific to the instrument that performed the observations, as described in the previous sections. The log-likelihood is used for finding best-fit parameters and to obtain the joint posterior distribution of all astrophysical and instrumental model parameters using the emcee package \citep{foreman-mackey_emcee:_2013}. All parameters have flat priors and initial values for the joint MCMC fit are set to the best-fitting parameters from the fit to the individual transit light curves. The median and 1$\sigma$ values of the orbital parameters and transit depths in the \textit{K2} and \textit{Spitzer} bandpasses are quoted in Tables \ref{tab:orb_para} and \ref{tab:trans_sp}, respectively. Within emcee, we use four walkers per free parameter and deem the chains converged once their lengths exceed at least 50 times the autocorrelation time computed using the new “Farad” method defined in the emcee 3.0.0 documentation. The latter is a more conservative approach compared to autocorrelation time defined in \citet{goodman_ensemble_2010}.

\begin{table}
\begin{tabular}{lcccc}
\hline
\hline
Instrument               &  Wavelength            &  Depth       & +1$\sigma$ &  -1$\sigma$ \\
                         &  ($\mu$m)              &  (ppm)       & (ppm)      &  (ppm) \\
\hline
\textit{K2}                       &  0.43 -- 0.89    &  2872.9     & 51.9    &  53.1 \\
\textit{HST}/WFC3                 &  1.12 -- 1.15    &  2916.2     & 27.9    &  26.8 \\
                         &  1.15 -- 1.18    &  2900.0     & 24.7    &  25.7 \\
                         &  1.18 -- 1.21    &  2872.5     & 24.5    &  24.6 \\
                         &  1.21 -- 1.24    &  2909.7     & 23.0    &  23.0 \\
                         &  1.24 -- 1.27    &  2909.2     & 24.5    &  24.0 \\
                         &  1.27 -- 1.30    &  2898.1     & 22.4    &  23.1 \\
                         &  1.30 -- 1.33    &  2896.9     & 22.8    &  22.7 \\
                         &  1.33 -- 1.36    &  2969.2     & 23.2    &  22.8 \\
                         &  1.36 -- 1.39    &  2954.5     & 22.8    &  22.3 \\
                         &  1.39 -- 1.42    &  2953.4     & 23.4    &  24.4 \\
                         &  1.42 -- 1.45    &  2970.9     & 23.0    &  22.2 \\
                         &  1.45 -- 1.48    &  2907.1     & 22.8    &  22.8 \\
                         &  1.48 -- 1.51    &  2934.7     & 23.3    &  22.5 \\
                         &  1.51 -- 1.54    &  2890.9     & 22.2    &  22.9 \\
                         &  1.54 -- 1.57    &  2868.6     & 24.7    &  25.0 \\
                         &  1.57 -- 1.60    &  2895.7     & 21.9    &  22.1 \\
                         &  1.60 -- 1.63    &  2878.2     & 23.5    &  24.5 \\
\textit{Spitzer}/IRAC Ch1         &  3.15 -- 3.94    &  2849.3     & 91.0    &  89.3 \\
\textit{Spitzer}/IRAC Ch2         &  3.96 -- 5.02    &  2882.2     & 93.6    &  91.6 \\
\hline
\hline
\end{tabular}
\caption{\label{tab:trans_sp} Optical/IR Transmission Spectrum of K2-18b}
\end{table}

\subsection{HST/WFC3 Transit spectroscopy} \label{sec:tr_specfit}

Following standard practice we use the results of the white-light-curve fitting to pre-correct the systematics in each spectroscopic light curve. This is possible because the ramp-like systematic variations in the \textit{HST}/WFC3 data are to first-order independent of wavelength. We test two methods for the correction: either dividing each of the spectroscopic time series by its corresponding best-fit systematics model from the white-light-curve analysis, or dividing it by the ratio of the white light curve to its best-fitting transit model. We find no significant difference between the two approaches for pre-correcting the light curves in terms of the derived parameters.

\begin{figure*}[t!]
\begin{center}
\includegraphics[width=0.8\linewidth]{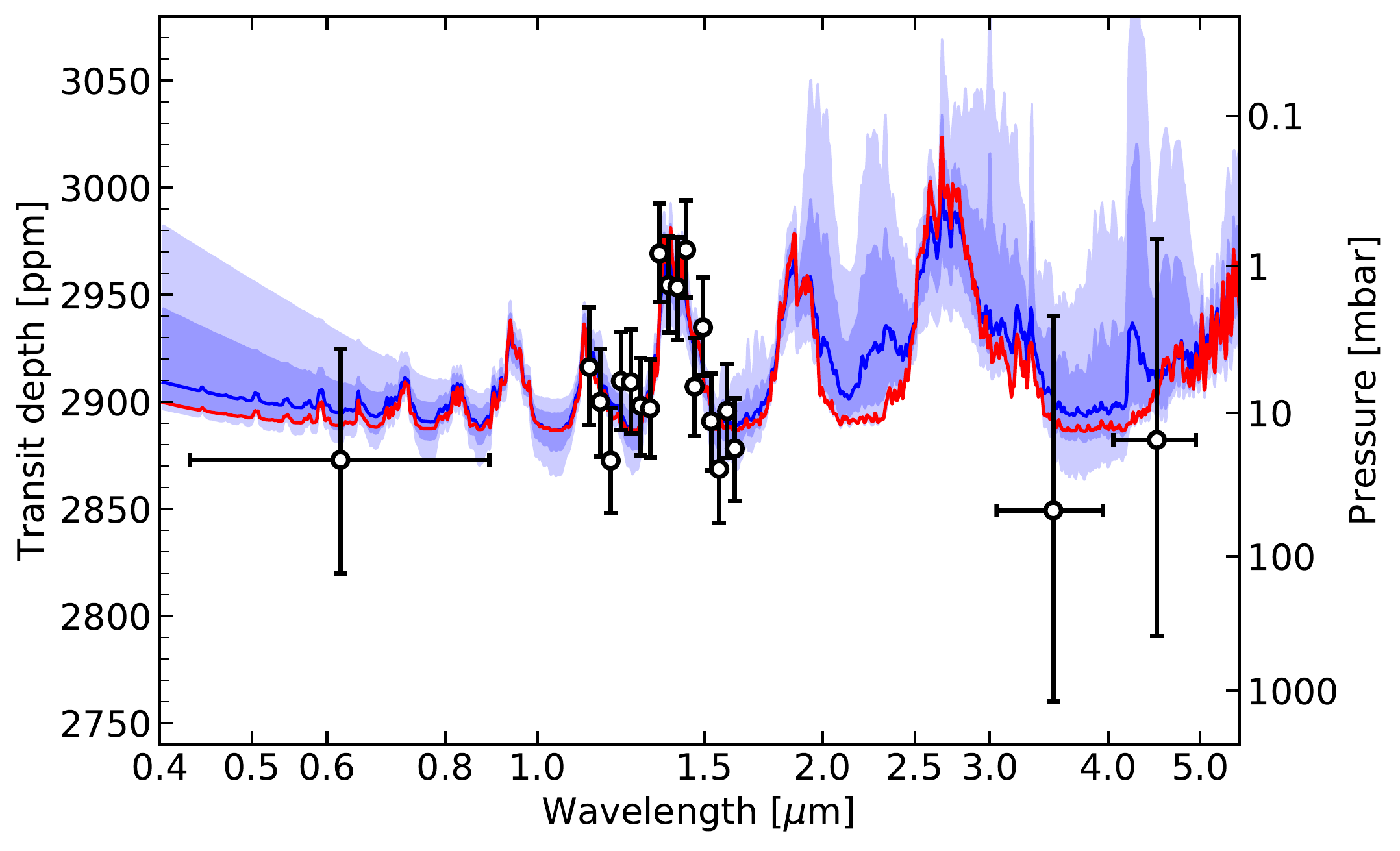}
\end{center}
\caption{Transmission spectrum of K2-18b computed from our global spectroscopic and broadband transit light-curve analysis (black points), and a random sampling of the model transmission spectra in the retrieval MCMC chain (blue). The shaded regions indicate 1$\sigma$ and 2$\sigma$ credible intervals in the retrieved spectrum (medium and light blue, respectively), relative to the median fit (dark blue line) and the overall best-fitting model (red). The main feature of the transmission spectrum is the prominent increase in transit depth within the 1.4~$\mu$m vibrational band of water vapor covered by the \textit{HST}/WFC3 data. The \textit{K2} data point is plotted at visible wavelengths and the \textit{Spitzer}/IRAC measurements are indicated at 3.6 and 4.5~$\mu$m. The secondary vertical axis on the right indicates the atmospheric pressure for the best-fitting model.}
\label{fig:spectrum}
\end{figure*}

In each spectroscopic channel, we then perform a joint MCMC fit to the eight pre-corrected spectroscopic light curves from the eight \textit{HST}/WFC3 transits. We keep the parameters $a/R_\star$, $b$, $T_0$, and $P$ fixed to the best-fitting values from the white-light-curve fit and fit a simplified systematics model that detrends linearly against the $x$ position of the trapezoidal spectral trace on the detector. We obtain consistent transmission spectra when fitting either a single linear limb-darkening coefficient at each wavelength or two coefficients for a quadratic limb-darkening profile. A consistent transmission spectrum is also obtained when using pre-computed limb-darkening coefficients from stellar atmosphere models (see \citet{lothringer_hst/stis_2018} and \citet{benneke_sub-neptune_2019} for details). The resulting WFC3 transit depths for the latter are quoted in Table \ref{tab:trans_sp} and depicted in Figure \ref{fig:spectrum} together with the \textit{K2} and \textit{Spitzer} transit depths. A typical spectral-light-curve fit for one wavelength bin is depicted in Figure \ref{fig:HST}. 

\subsection{Stellar Activity}

K2-18 is a moderately active M dwarf with a measured rotation period of approximately 39 days and corresponding photometric variability of 8 mmag in the \textit{Kepler} bandpass \citep{cloutier_characterization_2017}, 9 mmag in $B$-band, and 7 mmag in $R$-band \citep{sarkis_carmenes_2018}.  We see no evidence for spot occultations in any of our transit light curves, and when comparing the measured transit depth across multiple visits in the same bandpass we obtain consistent values, implying that the observed transit depths are relatively insensitive to variable spot coverage on the star. 

Still, to further rule out a stellar origin, we attempt to fit the transit spectrum with a stellar contamination model. Following \citet{rackham_transit_2018} we consider the hypothesis that the observed transit depth variations are due to inhomogeneities on the stellar surface. The observed wavelength-dependent transit depths $D_{\lambda,obs}$ are modeled as:

\begin{equation}
    D_{\lambda,obs} = \frac{D_\lambda}{1-f_{spot}\left(1-\frac{F_{\lambda,spot}}{F_{\lambda,phot}}\right) - f_{fac}\left(1-\frac{F_{\lambda,fac}}{F_{\lambda,phot}}\right)}
\end{equation}

\noindent where $D_{\lambda}$ is the geometric radius ratio $(R_p/R_\star)^2$, assumed here to be constant with wavelength ($=D$), $f_\mathrm{spot}$ and $f_\mathrm{fac}$ are the spot and faculae covering fractions, and the $F_{\lambda}$ refer to bandpass-integrated PHOENIX model spectra of the photosphere, spots, and faculae \citep{hauschildt_nextgen_1999}. Using K2-18b stellar parameters (Table \ref{tab:orb_para}), we explore a wide range of possible stellar contamination spectra for M2--M3 stars with spot and faculae covering fractions consistent with 1\% $I$-band variability. Based on \citet{rackham_transit_2018}, we then explore scenarios with giant spots and solar-like spots as well as with and without faculae, resulting in spot fractions of up to 20\%. However, none of those models can explain the amplitude of the observed transit spectrum. The amplitude of resulting transit depth variations due to the stellar inhomogeneities are up to 3--8\,parts-per-million (ppm), about an order of magnitude smaller than the observed 80--90\,ppm transit depth variations in the observed WFC3 spectrum (Figure \ref{fig:spectrum}). Finally, we also explore the most extreme scenarios where the spot and faculae covering fractions can be as high as 100\%, but even those stellar  inhomogeneity models fail to explain the amplitude of the observed transit depth variation. They deliver an absolute maximum of 20\,ppm at $1.4\,\mu$m, which still only corresponds to less than a quarter of the transit depth variation in the observations. We conclude that stellar inhomogeneities and activity cannot explain the measured transmission spectrum.

\section{Atmospheric Modeling}\label{sec:atmosModel}

We compute quantitative constraints on the atmosphere of K2-18b using the SCARLET atmospheric retrieval framework \citep{benneke_atmospheric_2012,benneke_how_2013,knutson_featureless_2014,kreidberg_clouds_2014,benneke_strict_2015,benneke_sub-neptune_2019}. To be as independent of model assumption as possible, we employ the ``free retrieval'' mode, which parameterizes the mole fractions of the molecular gases, the pressure of the cloud deck, and the atmospheric temperature as free fitting parameters. SCARLET then determines their posterior constraints by combining the atmospheric forward model with a Bayesian MCMC analysis. 

\begin{figure}[]
\begin{center}
\includegraphics[width=\linewidth]{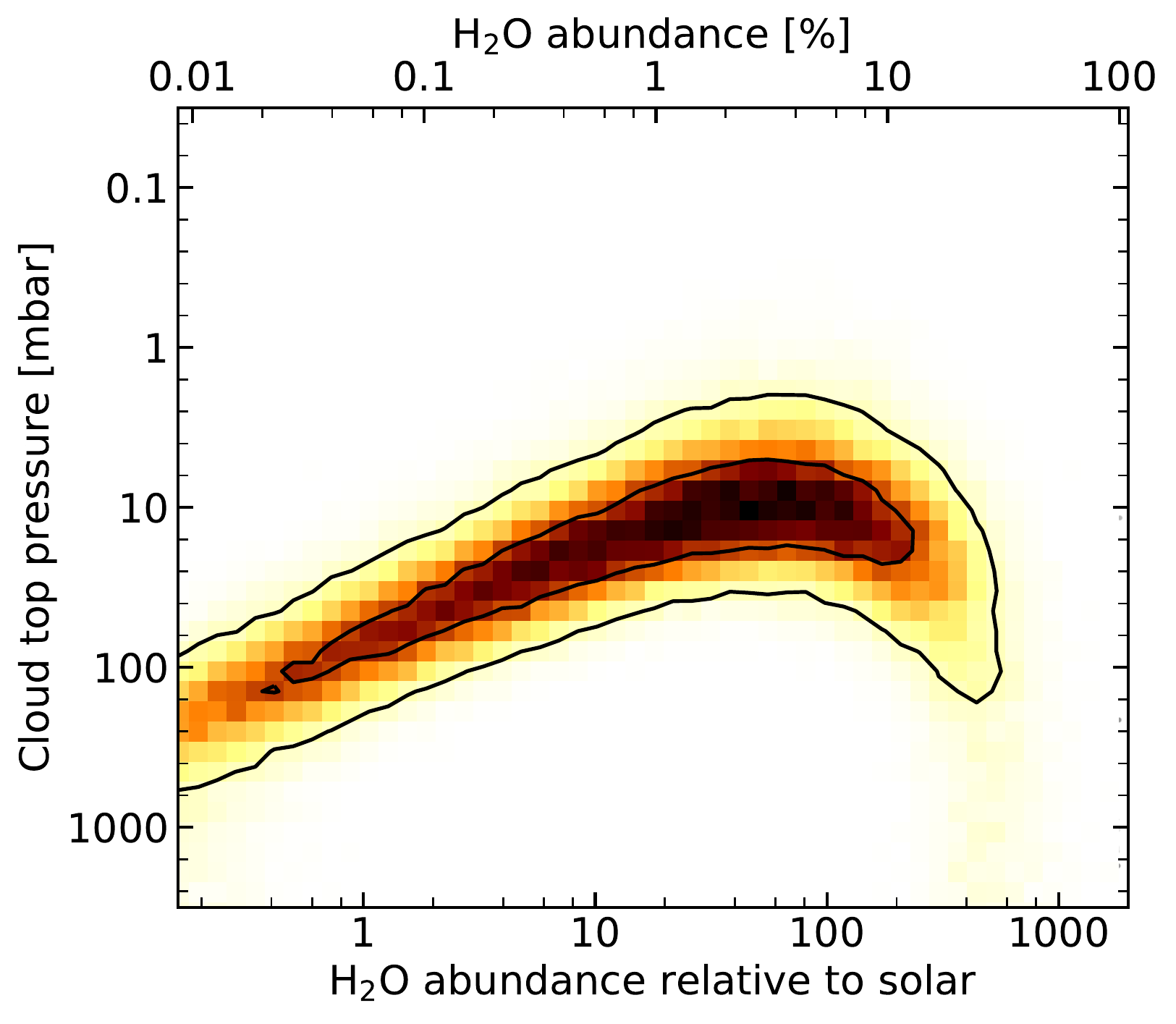}
\end{center}
\caption{Joint constraints on the water abundance vs. the cloud top pressure. The colored shading indicates the normalized probability density as a function of the water vapor mole fraction above the clouds and cloud top pressure derived using our Bayesian atmosphere retrieval framework. The black contours show the 1$\sigma$ and 2$\sigma$ regions. Absolute water vapor mole fraction is indicated at the top axis; the water abundance relative to a solar composition atmosphere is shown at the bottom.} \label{fig:cloudmetal}
\end{figure}

To evaluate the likelihood for a particular set of parameters, the atmospheric forward model first computes a model atmosphere in hydrostatic equilibrium, then determines the opacities of molecules at each layer, and finally computes the transmission spectrum. Beyond H$_2$/He, our model allows for H$_2$O, CH$_4$, CO, CO$_2$, NH$_3$, HCN, and N$_2$ with a log-uniform prior for mixing ratios between $10^{-10}$ and 1. We find that only H$_2$O is required by the data and that including the other molecules has virtually no impact on the best fit to the data. Following \citet{benneke_atmospheric_2012, benneke_how_2013}, we also include a cloud deck at a freely parameterized cloud top pressure with a log-uniform prior between $0.1\,$mbar and $10\,$bar. The cloud deck is assumed to be opaque to grazing light beams below the cloud top pressure as would occur for large droplets. We also explored a more complex three-parameter Mie-scattering cloud description as introduced in \citet{benneke_sub-neptune_2019}; however, we find no significant improvement in the fit to the observed transmission spectrum compared to gray clouds. Our atmospheric temperature is parameterized using a single free parameter for the mid-atmosphere probed by the observations because low-resolution transmission spectra are largely insensitive to the exact vertical temperature structure \citep{benneke_atmospheric_2012}. We also considered a five-parameter analytic model \citep{parmentier_non-grey_2014, benneke_sub-neptune_2019}, but no gradients in the temperature-pressure profile can be constrained.

High-resolution synthetic transmission spectra are computed using line-by-line radiative transfer, which are integrated over the appropriate instrument response functions to obtain synthetic observations to be compared to the observations. Sufficient wavelength resolution in the synthetic spectra is ensured by repeatedly verifying that the likelihood for a given model is not significantly affected by the finite wavelength resolution ($\Delta\chi^2<0.001$). Reference models are computed at $\lambda/\Delta\lambda=250000$. Our final retrieval run computes all atmospheric models at $\lambda/\Delta\lambda=30000$. 

We perform all retrieval analyses with 100 walkers using log-uniform priors on all parameters. Following best practice for emcee \citep{foreman-mackey_emcee:_2013}, our formal convergence test is based on the integrated autocorrelation time \citep{goodman_ensemble_2010}, but we run the chains well beyond formal convergence to obtain smooth posterior distribution contours. We double-check that the number of iterations of our 100 walker chain exceeds at least 50 times the autocorrelation time computed based on \citet{goodman_ensemble_2010} as well as the autocorrelation time computed computed based on the more conservative “Farad” method in the emcee 3.0.0 documentation.

\section{Results}\label{sec:resu}

Our transmission spectrum of K2-18b reveals a slightly attenuated but statistically significant water absorption feature at $1.4\,\mu$m in our \textit{HST}/WFC3 data (Figure \ref{fig:spectrum}). The water absorption is detected in multiple neighboring spectroscopic channels covering the 1.4 $\mu$m water band and protrudes over an otherwise relatively flat visible to near-IR transmission spectrum. Quantitatively, retrieval models that include molecular absorption by water are favored by the Bayesian factor at 459:1 (3.93$\sigma$). We evaluate the Bayes factor by comparing the Bayes evidence for models with and without water opacity \citep{benneke_how_2013}. We also evaluate the AIC of our model in comparison to the simple flat line model with $R_P/R_\star$ as the only free parameter and find a strong preference of $\Delta$AIC=18.1 in favor of the model with water absorption. 

Our retrieval modeling also shows that the data are best matched by a hydrogen-dominated atmosphere with water vapor absorbing above clouds in the mid-atmosphere (Figure \ref{fig:cloudmetal}). Water abundances between 0.033\% and 8.9\% are most consistent with the data at 1$\sigma$, corresponding to 0.6 times and 162 times the value for a solar abundance atmosphere, respectively. The clouds become optically thick below the 7.7--139 mbar level. The water abundance and cloud top pressure are strongly correlated because models with similar water vapor column density above the cloud deck result in similar transmission spectra \citep[see also][]{benneke_how_2013}. For the lowest water abundances, pressure levels up to a few hundred millibars are probed by our observations because K2-18b's radius is relatively small and the low temperature results in a relatively small atmospheric scale height. Both effects reduce the path lengths for starlight grazing through the atmosphere compared to most previously studied exoplanets. 

Cloud-free models with water mixing ratios greater than several hundred times solar are substantially disfavored by our combined data set indicating that K2-18b hosts an atmosphere rich in hydrogen and helium. This is consistent with our new measurement of the planet density  (Section 3.1) which supports the presence of a thick hydrogen-rich gas envelope \citep{lopez_understanding_2014}. Quantitatively, the retrieval model with clouds in the mid-atmosphere is favored by at least 8.3:1 (2.6$\sigma$) in the Bayesian evidence over retrieval models without clouds. The significance rises further if extreme subsolar O/H ratios for K2-18b are not considered. We conclude that it is crucial to report the water abundance constraints from retrieval models that consider clouds. Cloud-free retrieval models would misleadingly favor extremely high water mole fractions and narrower constraints.

Finally, while no molecular species other than H$_2$O is directly detected in our observations of K2-18b, we can still infer useful upper limits on their mole fractions. The two dominant effects of additional molecules are the introduction of additional wavelength-dependent opacities and, at high mole fraction, the increase of the overall atmospheric mean molecular weight \citep{benneke_atmospheric_2012}. The latter, in turn, dampens the amplitudes of all molecular absorption features in the transmission spectrum. The upper limit on the spectrally inactive gas N$_2$, for example, is 10.9\% at 2$\sigma$ (97.5\%) because above that level the atmospheric mean molecular mass would be sufficiently increased to not allow for the amplitude of the detected $1.4\,\mu$m H$_2$O absorption feature. The other molecular gases (CO, CO$_2$, NH$_3$, and CH$_4$) are similarly constrained to 7.45\%, 2.4\%, 13.5\%, and 0.248\% at 2$\sigma$, respectively. Molecular opacities also played an important role for these constraints because CO, CO$_2$, NH$_3$, and CH$_4$ can result in substantial opacities within the \textit{HST}/WFC3 and/or \textit{Spitzer} bandpasses. 

\begin{figure}[tb]
\begin{center}
\includegraphics[width=\linewidth]{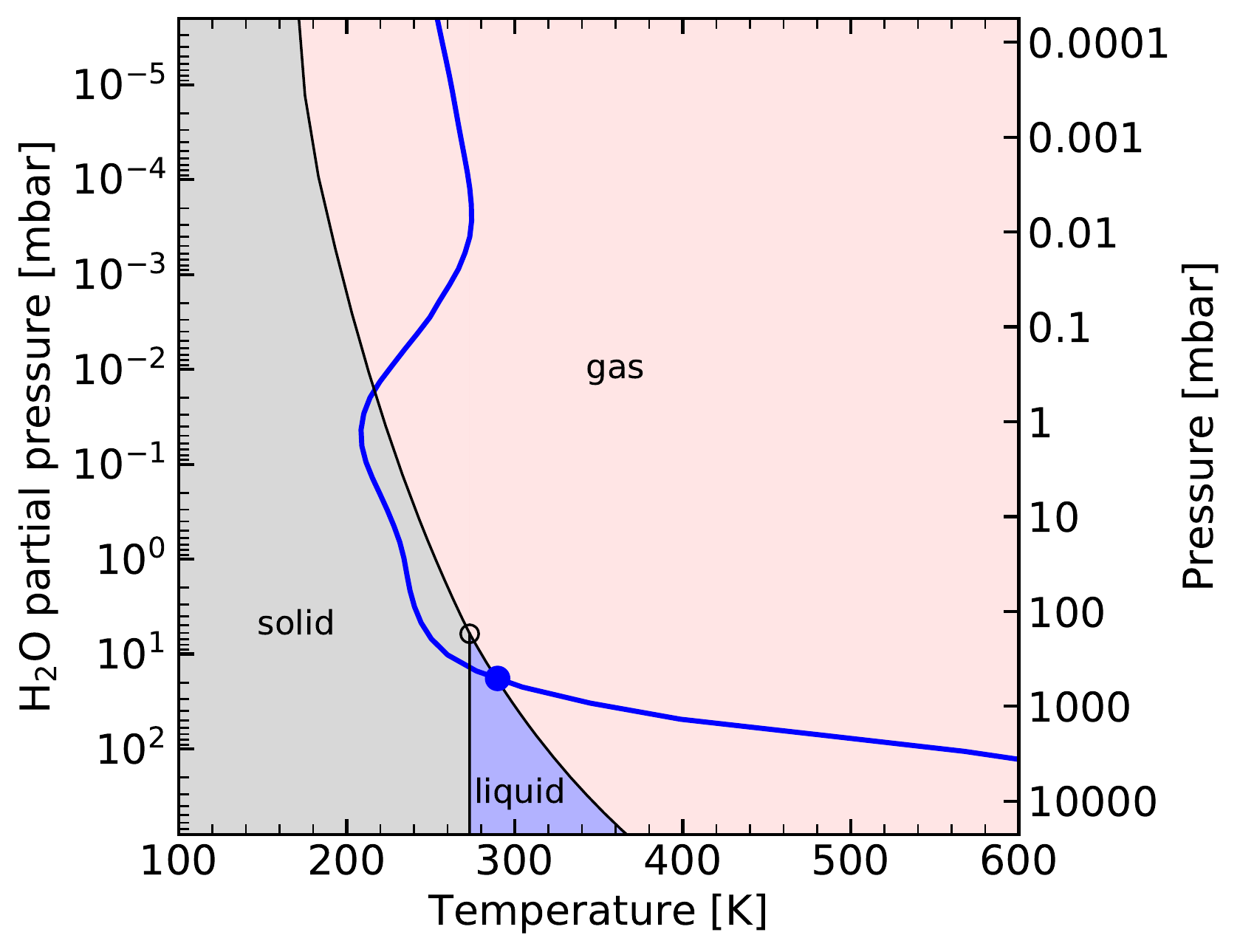}
\end{center}
\caption{Self-consistent temperature-pressure profile for the retrieved best-fitting molecular composition of K2-18b in comparison to the water phase diagram. The blue curve shows the vertical temperature profile for an Earth-like Bond albedo (A=0.3). The partial pressure of water is shown on the left; the total atmospheric pressure is shown on the right. The temperature profile crosses the liquid area of the water phase diagram (blue circle). Clouds that form in this regime form liquid water droplets and possibly result in liquid water rain. Icy clouds can possibly form at higher altitude as well. Note that the pressure axis is inverted as is commonly done for atmospheric temperature profiles.} \label{fig:WaterCond}
\end{figure}

\section{Discussion and Conclusions}\label{sec:disc}

The simultaneous detection of water vapor and clouds in the mid-atmosphere of K2-18b is particularly intriguing because K2-18b's stellar insolation and atmospheric temperature are sufficiently low for water vapor to condense. Water clouds are therefore a very plausible explanation for the detected clouds. To demonstrate the plausibility of liquid droplet formation, we explore self-consistent temperature structures for atmospheric
compositions consistent with the observations (Figure \ref{fig:WaterCond}). We computed a set of fiducial temperature-pressure profiles using the implementation described in \citet{benneke_sub-neptune_2019} and \citet{morley_quantitatively_2013}. Both models iteratively solve the radiative-convective heat transport. The models predict the onset of liquid condensation between approximately 10 and 1000 mbar at locations where the water vapor is supersaturated. The fact that the best-fitting retrieval models are obtained for cloud top pressures in the same range further supports the scenario of liquid water clouds (Figure \ref{fig:cloudmetal}).

The water vapor mixing ratio retrieved from our spectroscopic observations should be regarded as the mole fraction of water vapor in the gaseous atmosphere above the clouds. Because water condenses into clouds, this presents a lower limit on the overall water fraction in the atmosphere K2-18b. The water abundance in the upper atmosphere is set by a complex interplay between upward transport through vertical mixing and homogeneous/heterogeneous condensation and coagulation processes. Detailed modeling of these cloud formation processes will be presented in a follow-up study. 

The detection of water vapor makes K2-18b a key target for more detailed follow-up studies with the upcoming \emph{James Webb Space Telescope} (\emph{JWST}). Unlike any other temperate and low-mass planet, we now know that K2-18b shows evidence for atmospheric water vapor and is amenable to characterization via transmission spectroscopy. Our result suggests K2-18b orbits far enough away from its moderately active host star to avoid high levels of haze production, while also remaining cool enough that cloud species like KCl and ZnS that may exist in warmer atmospheres do not form. This may indicate that cooler planets like K2-18b occupy a sought after niche in the low-mass planet parameter space that is accessible to spectroscopic characterization with \emph{JWST}. \emph{JWST}'s wavelength coverage will extend from 0.55~$\mu$m to the thermal infrared, where many other molecular species like CH$_4$, CO, CO$_2$, and NH$_3$ can be probed directly. The higher precision and spectral resolution obtained from repeated \emph{JWST} transit observations will allow us to better constrain K2-18b's atmospheric composition and cloud properties, and potentially even look for biomarkers in the gas envelope of a habitable-zone exoplanet \citep[e.g.,][]{seager_biosignature_2013}.

\acknowledgements

This work is based on observations with the NASA/ESA HST, obtained at the Space Telescope Science Institute (STScI) operated by AURA, Inc. We received support for the analysis by NASA through grants under the HST-GO-13665 and HST-GO-14682 programs (PI Benneke). This work is also based in part on observations made with the \textit{Spitzer Space Telescope}, which is operated by the Jet Propulsion Laboratory, California Institute of Technology under a contract with NASA (PIs Benneke and Werner). B.B. further acknowledges financial supported by the Natural Sciences and Engineering Research Council (NSERC) of Canada and the Fond de Recherche Qu\'{e}b\'{e}cois—Nature et Technologie (FRQNT; Qu\'{e}bec). D.D. acknowledges support provided by NASA through Hubble Fellowship grant HST-HF2-51372.001-A awarded by the Space Telescope Science Institute, which is operated by the Association of Universities for Research in Astronomy, Inc., for NASA, under contract NAS5-26555.

\bibliography{main}



\end{document}